\documentclass[twocolumn,showpacs,superscriptaddress,aps,prl]{revtex4}
\usepackage{epsf,psfig}
\begin{document}
\def\pT{\mbox{$p_T$}}
\def\mpT{$\langle p_{T} \rangle\,$ }
\topmargin=0.0in
\title{

Pseudorapidity Asymmetry and Centrality Dependence of Charged Hadron
Spectra in d+Au Collisions at $\sqrt{s_{NN}}=200$ GeV}

\author{J.~Adams}\affiliation{University of Birmingham, Birmingham, United Kingdom}
\author{M.M.~Aggarwal}\affiliation{Panjab University, Chandigarh 160014, India}
\author{Z.~Ahammed}\affiliation{Variable Energy Cyclotron Centre, Kolkata 700064, India}
\author{J.~Amonett}\affiliation{Kent State University, Kent, Ohio 44242}
\author{B.D.~Anderson}\affiliation{Kent State University, Kent, Ohio 44242}
\author{D.~Arkhipkin}\affiliation{Particle Physics Laboratory (JINR), Dubna, Russia}
\author{G.S.~Averichev}\affiliation{Laboratory for High Energy (JINR), Dubna, Russia}
\author{S.K.~Badyal}\affiliation{University of Jammu, Jammu 180001, India}
\author{Y.~Bai}\affiliation{NIKHEF, Amsterdam, The Netherlands}
\author{J.~Balewski}\affiliation{Indiana University, Bloomington, Indiana 47408}
% ...

\collaboration{STAR Collaboration}\noaffiliation

\date{\today}
\begin{abstract}
The pseudorapidity asymmetry and centrality dependence of charged
hadron spectra in d+Au collisions at $\sqrt{s_{NN}}=200$ GeV are
presented. The charged particle density at mid-rapidity, its
pseudorapidity asymmetry and centrality dependence are reasonably
reproduced by a Multi-Phase Transport model, by HIJING, and by the latest calculations
in a saturation model.  Ratios of transverse momentum spectra between backward and forward 
pseudorapidity are above unity for \pT~below 5 GeV/$c$.  The
ratio of central to peripheral spectra in d+Au collisions shows
enhancement at 2 $<$ \pT $<$ 6 GeV/$c$, with a larger effect at
backward rapidity than forward rapidity.  Our measurements are in
qualitative agreement with gluon saturation and in contrast to
calculations based on incoherent multiple partonic scatterings.
\end{abstract}
\maketitle
Soft and hard scattering processes have distinctive rapidity and
centrality dependences in the context of particle production in
d(p)+Au collisions.  Models based on the Color Glass
Condensate~\cite{dima,cgc}, HIJING~\cite{wang}, and Multi-Phase Transport
(AMPT)~\cite{ziwei} predict specific pseudorapidity ($\eta$) and
centrality dependence of produced particle density which can be
directly compared to experimental measurements. The Cronin
effect~\cite{Cronin}---the enhancement of particle yield at
intermediate transverse momentum (\pT) with respect to binary
collision scaling---has also been observed in d+Au collisions at
RHIC~\cite{STARdAu200,PHENIXdAu200,PHOBOSdAu200,BRAHMSdAu200,starpiddAu}.
For partonic processes such as the dominant $g+g$ and $q+g$
scatterings, the particle rapidity distribution can be evaluated in a
pQCD-inspired framework that depends on the parton distribution
functions and the underlying dynamics. For example, calculations of
the Cronin effect based on incoherent initial multiple partonic scatterings
and independent fragmentation~\cite{wang} predict a unique rapidity
asymmetry of particle production in d+Au collisions, where the
backward-to-forward (negative rapidity (Au) to positive rapidity (d))
particle ratio is greater than unity at low \pT, goes below unity at
intermediate \pT, and approaches unity again at high \pT. The
amplitude of the theoretical backward-to-forward particle ratios
depends on the nuclear shadowing~\cite{wang}. Calculations of shadowing alone, based on Regge theory and hard
diffraction~\cite{strikman}, are fairly successful in describing the observed
suppression of particle production at forward rapidity in d+Au
collisions~\cite{vogt}.   The calculation in Ref.~\cite{vogt} considers the
spatial dependence of the shadowing, leading to an impact parameter
dependence that goes beyond the simple geometrical scaling.
Calculations in a gluon saturation model~\cite{KKT2}
predict a backward-to-forward particle ratio that is opposite to
the predictions based on incoherent multiple partonic scatterings. In this
approach, the particle production is related to the high gluon density
in the nucleus (nucleon).  The asymmetry is greater than unity in the
range of transverse momenta determined by the values of the saturation
scale $Q_s(y)$ and the geometrical scale $Q_s^2(y)/Q_{s, min}$, where
$Q_{s, min}$ is at the onset of the gluon saturation. Recently, the
quark recombination model was used to explain the Cronin effect as a
final state effect~\cite{finalstate}, implying a backward-to-forward
particle ratio markedly different from that of the QCD-inspired
formulation in~\cite{wang} and similar to the predictions by a saturation model~\cite{KKT2}. In this approach, the enhancement of
particle production at intermediate \pT~is an extension from low
\pT~due to the thermal parton and shower parton
recombination~\cite{finalstate}.

The suppression of high transverse momentum particles in
central Au+Au collisions at RHIC can be described by both final state
and initial state effects, such as jet quenching calculations that
assume parton energy loss via gluon
bremsstrahlung~\cite{Eloss1,Eloss2} or gluon saturation~\cite{KLM}.
The measurement of particle production at mid-rapidity from d+Au
collisions at
RHIC~\cite{STARdAu200,PHENIXdAu200,PHOBOSdAu200,BRAHMSdAu200} favors
the scenario that the suppression of high-\pT~particles is primarily
due to the final state interactions, i.e., processes after the hard
partonic scattering.  The quantitative features of high-\pT~particle
production in Au+Au collisions can be described by models that
incorporate a combination of physical effects such as the Cronin
effect, nuclear shadowing~\cite{shadow}, and parton energy
loss~\cite{Eloss1,Eloss2}. The Cronin effect and shadowing can be
investigated in d(p)+Au collisions. The magnitude of these nuclear
effects on particle production has a geometrical dependence due to
the nuclear density distribution.  The particle production in d(p)+Au collisions at
different rapidities also reflects the dynamics of nuclear and
Bjorken-$x$ dependence of these effects.  Therefore, the centrality,
pseudorapidity and \pT~dependence of particle production in d(p)+Au
collisions provides an essential baseline for understanding the
underlying phenomena in Au+Au collisions.

We present inclusive \pT~spectra of charged hadrons over an $\eta$
range of $-$1(Au-side) to $+$1(d-side) in d+Au collisions at
$\sqrt{s_{NN}}=200$ GeV with several collision centrality selections.
For these measurements, the STAR Time-Projection Chamber
(TPC)~\cite{TPC} provided tracking of charged hadrons.  The minimum
bias trigger was defined by requiring that at least one beam-rapidity
neutron impinge on the Zero Degree Calorimeter~\cite{ZDC} in the Au beam
direction. The measured minimum bias cross section amounts to
95$\pm$3\% of the total d+Au geometric cross section.  Charged
particle multiplicity within $-$3.8 $<$ $\eta$ $<$ $-$2.8 was measured
by the Forward TPC~\cite{FTPC} in the Au beam direction and served as
the basis for our d+Au centrality tagging scheme, as described
in~\cite{STARdAu200}.  The d+Au centrality definition consists of
three event centrality classes; the 0-20, 20-40 and 40-100 percentiles
of the total d+Au cross section.  A separate centrality tag, which
requires that a single neutron impinge on the Zero Degree Calorimeter in
the deuteron beam direction (ZDC-$d$), was also used.  Our analysis
was restricted to events with a primary vertex within 50 cm of the
center of the TPC along the beam direction.  This yielded a data set
of $9.5\times10^6$ minimum bias events.  Only tracks (with at
least 15 measured points) with a projected distance of closest
approach to the event primary vertex of less than 3 cm were used in
the analysis.
\begin{figure}
\centering\mbox{
\psfig{figure=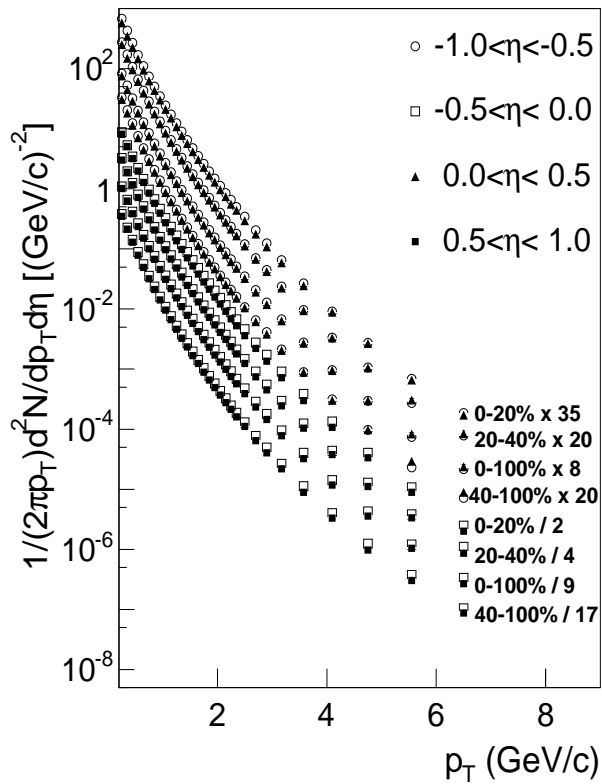,width=8cm,height=10.5cm}}
  \caption{
    The \pT~spectra of charged hadrons. From the top, the
    open circles correspond to the 0-20\%,
    20-40\%, minimum bias, and 40-100\% centralities in $-$1.0 $<$
    $\eta$ $<$ $-$0.5. 
    Similarly, the solid triangles, open
    squares, and solid squares correspond to \pT~spectra in 0.0 $<$
    $\eta$ $<$ 0.5, $-$0.5 $<$ $\eta$ $<$ 0.0, and 0.5 $<$ $\eta$ $<$ 1.0,
    respectively. Spectra have been scaled by the factors indicated in the figure.}
  \label{ptspectra_etacuts}
\end{figure}

Acceptance and TPC tracking efficiency corrections in various
pseudorapidity regions and centrality classes were obtained by
embedding simulated data into a real data sample.  In the region of
$|\eta|<$ 0.5, the tracking efficiency and acceptance above \pT~= 2.0
GeV/$c$ were observed to reach a plateau of about 90\% for all
centrality classes. Efficiency corrections using filtered
HIJING~\cite{HIJING}---HIJING events in a GEANT simulation of the
detector---were also used; a maximum difference between HIJING and
embedded data of about 3\% was observed. Background due to weak decay
products was accounted for using filtered HIJING.  For the 0-20\% most
central events, the contaminating signals are estimated at less than
18\% for \pT$<$1.0 GeV/$c$, and for the 40-100\% most peripheral events
this was observed to be less than 12\%. The background exponentially
decreases, and above \pT~= 1.0 GeV/$c$, the background is approximately
4\%, exhibiting no strong dependence on centrality or pseudorapidity.
A net uncertainty of 6\% in the analysis corrections was determined by
adding the efficiency and background correction uncertainties in
quadrature.  

\begin{figure}
\centering\mbox{
\psfig{figure=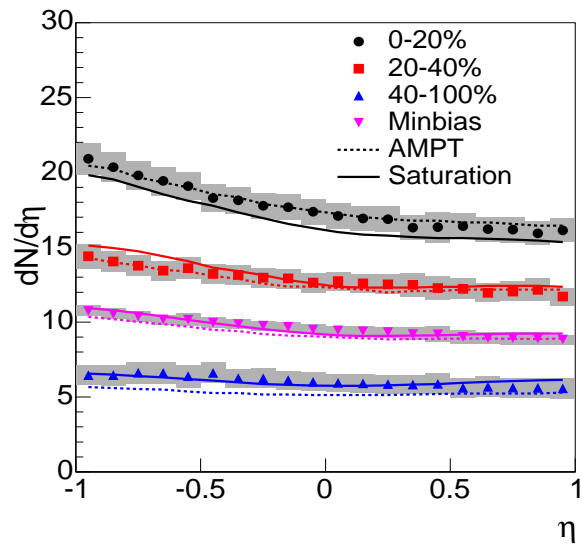,width=8cm,height=7.2cm}}
  \caption{(Color online) The pseudorapidity dependence of charged particle densities
    for various centrality classes. Particle tracking efficiency and
    background corrections were carried out for each pseudorapidity
    bin ($\Delta$$\eta$=0.1).  The point-to-point systematic
    uncertainties shown for each distribution (indicated by bands) are
    the quadratic sum of the efficiency and background correction
    uncertainties; statistical uncertainties are negligible.  The
    results of AMPT (with default parameters) and parton saturation
    are indicated by the dashed and solid lines, respectively.}
  \label{eta_dist}
\end{figure}

The transverse momentum spectra of primary charged hadrons
for various pseudorapidity regions are shown in
Fig.~\ref{ptspectra_etacuts} for the 0-20\%, 20-40\%, 40-100\%
centrality selections, and for minimum bias events.  In the region of
0.2 $<$ \pT $<$ 2.0 GeV/$c$, the charged hadron spectra were fitted
with a power law function:
\begin{equation}
{{d^{2}N}\over{p_{T}dp_{T}d\eta}}={A\over{(1+{p_{T}/{p_{0}}})^{n}}}.
\end{equation}
\noindent
The integrated charged hadron multiplicity per unit of pseudorapidity
$dN/d\eta$ was obtained by summing up the measured yields in the
covered momentum range and using the power-law function for
extrapolation to \pT~= 0 GeV/c.  Fig.~\ref{eta_dist} shows the
pseudorapidity dependence of charged particle densities for various
centrality classes.  Calculations based on the ideas of gluon
saturation~\cite{dima} in the Color Glass Condensate as well as the
predictions of AMPT~\cite{ziwei} are also shown. Both models predict
a similar pseudorapidity dependence of particle yields.  It should be
noted that the pseudorapidity and centrality dependence of charged
particle yields generated by HIJING~\cite{HIJING} (without shadowing)
are nearly identical to the AMPT results at mid-rapidity. There is a
clear increase in the asymmetry of charged particle densities as a
function of increasing centrality: a prominent pseudorapidity
dependence is observed for the 0-20\% most central collisions, while
peripheral collisions between gold nuclei and deuterons are akin to
symmetric p+p collisions. The predictions of the gluon saturation
model and AMPT are in good overall agreement with the data. 

\begin{figure}
\centering\mbox{
\psfig{figure=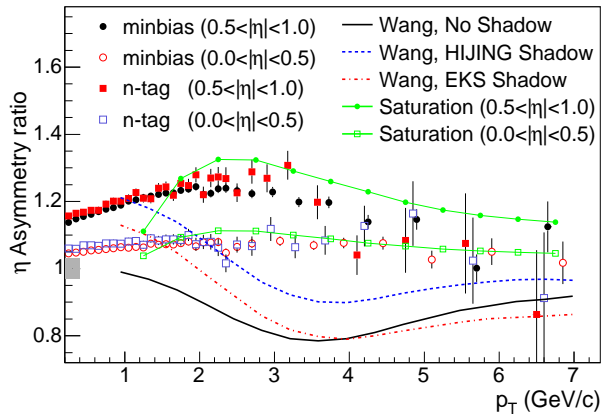,width=8cm,height=5.5cm}}
  \caption{
    (Color online) The ratio of charged hadron spectra in the backward rapidity to forward rapidity
    region for minimum bias and ZDC-$d$ neutron-tagged events.
    Calculations based on pQCD~\cite{wang} ($y$=$-$1/$y$=1) for minimum bias
    events are also shown for cases with 
    no shadowing (solid curve), HIJING shadowing (dashed curve), and 
    EKS shadowing (dot-dashed curve). Calculations in a gluon
    saturation model~\cite{KKT2} 
    for minimum bias events are shown for 0.5 $<$ $|\eta|$ $<$ 1.0 (filled circles with solid line) and for 
    0.0 $<$ $|\eta|$ $<$ 0.5 (open squares with solid line).} 
  \label{ptratios}
\end{figure}
\begin{figure}
\centering\mbox{
\psfig{figure=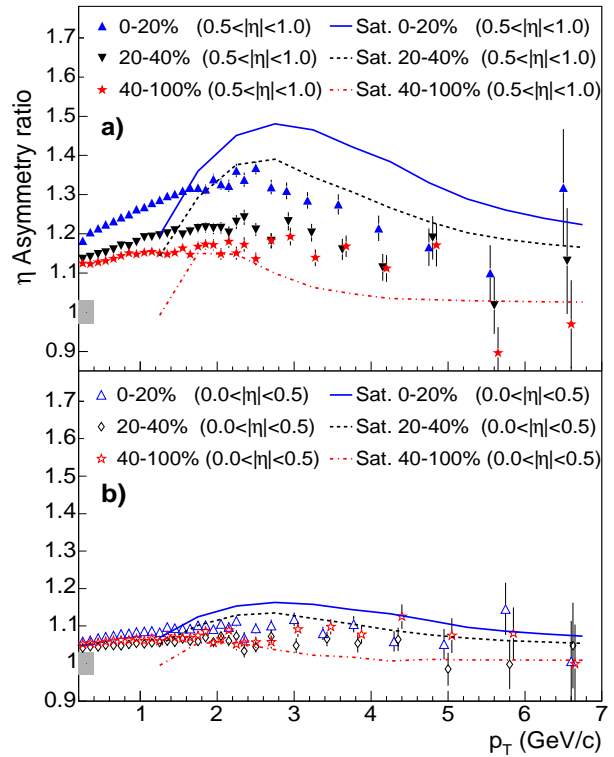,width=8cm,height=10cm}}
  \caption{(Color online) a) The centrality dependence of the ratio of charged
    hadron spectra in backward rapidity to forward rapidity (0.5 $<$
    $|\eta|$ $<$ 1.0).  The gluon saturation model calculations
    are also shown for the 0-20\% (solid curve), 20-40\%
    (dashed curve), and 40-100\% (dot-dashed curve) centrality
    classes.  b) The centrality dependence of the ratio of charged
    hadron spectra in backward rapidity to forward rapidity (0.0 $<$
    $|\eta|$ $<$ 0.5).  The gluon saturation model calculations
    are also shown for the 0-20\% (solid curve), 20-40\%
    (dashed curve), and 40-100\% (dot-dashed curve) centrality
    classes.}
  \label{ptratios2}
\end{figure}
We define a measured asymmetry by taking ratios of inclusive backward
(Au-side) to forward (d-side) \pT~spectra.  Fig.~\ref{ptratios} shows
the \pT~dependence of the asymmetry for minimum bias and ZDC-$d$
neutron-tagged events.  The ratio was taken between the
$-$1.0$<\eta<$$-$0.5 and 0.5$<\eta<$1.0 as well as $-$0.5$<\eta<$0.0
and 0.0$<\eta<$0.5 regions.  An overall systematic uncertainty
(indicated by the band) of less than 3\% was assessed by taking the
corresponding ratios between inclusive spectra measured by STAR in p+p
collisions at the same energy, where an asymmetry is not expected to
be present. The ratio taken within $|\eta|<$ 0.5 is nearly constant in
\pT, with a maximum value of approximately 1.075.  This indicates that
there is a small disparity between the forward and backward regions
immediately around $\eta$ = 0.  The ratio taken at higher
pseudorapidity slowly increases with \pT~up to about \pT~= 2.5
GeV/$c$, attaining a value of approximately 1.25.  The ratio taken at
higher pseudorapidity approaches unity beyond \pT~= 5 GeV/$c$,
indicating the absence of nuclear effects at high \pT.  For the ZDC-$d$
neutron-tagged events, the ratio exhibits nearly the same
\pT~dependence as minimum bias events.  Fig.~\ref{ptratios2}a
illustrates the centrality dependence of the asymmetry in the region
of 0.5 $<$ $|\eta|$ $<$ 1.0.  The asymmetry becomes more prominent
with increasing centrality, reaching a factor of about 1.35 for the
most central events.  The asymmetry in the region of 0.0 $<$ $|\eta|$
$<$ 0.5, shown in Fig.~\ref{ptratios2}b, does not exhibit a strong
centrality and \pT~dependence. The neutron-tagged events have an average
number of binary collisions,
$\langle N_{bin} \rangle = 2.9 \pm 0.2$, well
below the $\langle N_{bin} \rangle = 7.5 \pm 0.4$
of the minimum bias data set.  The events in which a
single nucleon from the deuteron interacted with
the Au nucleus comprise approximately half of
the 40-100\% peripheral centrality class~\cite{STARdAu200}.  However,
Fig.~\ref{ptratios} shows that the $\eta$
asymmetry ratios for minimum bias and
neutron-tagged events are nearly identical.

Particle production at mid-rapidity in d+Au collisions may include
contributions from deuteron-side partons that have experienced
multiple scatterings while traversing the gold nucleus, and from
gold-side partons that may have been modified by nuclear effects.
Also shown in Fig.~\ref{ptratios} is the calculation of the asymmetry
in the incoherent multiple partonic scattering framework with various
nuclear shadowing parameterizations: no nuclear shadowing, the HIJING
shadowing~\cite{HIJINGshadow}, and the EKS shadowing~\cite{EKS}
parameterizations.  The ratio, taken for minimum bias spectra at $y$ =
$-$1 and $y$ = 1, is below unity at $p_T \sim$ 3-4 GeV/$c$ and is a
consequence of the increase in \pT~for partons from the deuteron
hemisphere.  Our measurements disagree with the theoretical
calculations~\cite{wang} and thus suggest that incoherent multiple
scattering of partons in the initial state alone cannot reproduce the
observed pseudorapidity asymmetry in the intermediate \pT~region.  By
the same token, the class of models that incorporate initial parton
scattering~\cite{wang,ziwei,Accardi}, though capable of reproducing
integrated observables such as charged particle yield asymmetries, may
not adequately reproduce the \pT~dependence of the asymmetry.  In
this respect, the \pT~dependence of the pseudorapidity asymmetry as
illustrated by the backward-to-forward ratio of charged hadron spectra
can serve as an important discriminator between models.

The minimum bias
gluon saturation results for the backward-to-forward ratio of charged
hadron spectra, also shown in Fig.~\ref{ptratios}, were obtained by
performing a calculation identical to the one in Ref.~\cite{KKT2} on
the basis of the method developed in~\cite{KKT1,KLM}.  In this
approach, the asymmetry is greater than unity in the range of
transverse momenta determined by the values of the saturation scale
$Q_s(y)$ and the geometrical scale $Q_s^2(y)/Q_{s, min}$.  The
calculated particle yield aysmmetry, evaluated over the same
pseudorapidity range as the data, is in qualitative agreement with our
observations.  The theoretical asymmetry exhibits a stronger
\pT~dependence than actually observed, overpredicting the magnitude of
the asymmetry at high pseudorapidities.  The centrality dependence of
the backward-to-forward particle yields in a saturation model, illustrated
in Fig.~\ref{ptratios2}a and Fig.~\ref{ptratios2}b, qualitatively
reproduces the observed centrality dependence.  Although the model
calculations fail to describe the data in detail, they show the same
trend of increasing asymmetry with increasing centrality. 
We note that some conventional models~\cite{vogt,Qiu-Vitev} are able to reproduce the suppression
of particle production at forward rapidity in d+Au collisions, which
was thought to be a unique feature of gluon
saturation~\cite{cgc,KKT2,BRAHMS}. It will be interesting to 
quantitatively compare our measurements with those calculations in the future.

It should be noted that a strong particle dependence in the nuclear
modification factor has been observed in this intermediate \pT~region
in both Au+Au~\cite{paulRcp} and d+Au
collisions~\cite{starpiddAu}. Collective partonic effects at the
hadron formation epoch such as parton coalescence or
recombination~\cite{reco1,reco2,reco3,reco4} have been proposed to
explain Au+Au results. The pseudorapidity asymmetry approaches unity
at a \pT~scale above 5 GeV/$c$, approximately the same \pT~scale above
which the particle dependence of the nuclear modification factor
disappears.  The idea of recombination was modified to explain the
Cronin effect and its particle dependence~\cite{finalstate} as a final
state effect.  In this approach, the enhancement of particle
production at intermediate \pT~is an extension from low \pT~due to the
thermal parton and shower parton recombination~\cite{finalstate},
qualitatively consistent with the measurements of the pseudorapidity
asymmetry as a function of \pT.  We should emphasize that the pseudorapidity
asymmetry is not likely to be solely due to the change of particle
composition.  In the recombination model, the shower and thermal
parton recombination not only enhances the baryon production, but also 
the meson production~\cite{finalstate}.  The pseudorapidity asymmetry of identified
pion spectra and its quantitative comparison to models are important for
further understanding of particle production at intermediate \pT.

\begin{figure}	
\centering\mbox{	
\psfig{figure=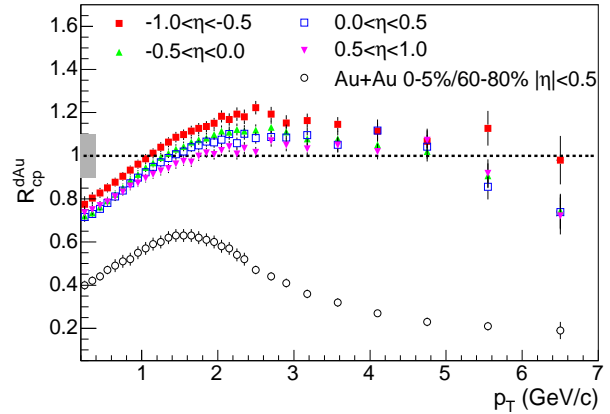,width=8cm,height=5.5cm}}
  \caption{
    (Color online) The ratio of central (0-20\%) to peripheral (40-100\%) spectra in d+Au collisions for
    various pseudorapidity regions and in Au+Au collisions at mid-rapidity.}
  \label{Rcp}
\end{figure}

Of similar interest is the ratio of d+Au central to
peripheral inclusive spectra
\begin{equation}
R_{CP}^{\rm{dAu}} =
\frac{(d^{2}N/dp_{T}d\eta/\langle
N_{\rm{bin}} \rangle)|_{\rm{central}}}{(d^{2}N/dp_{T}d\eta/\langle N_{\rm{bin}} \rangle)|_{\rm{periph}}},
\end{equation}     
\noindent
where $d^{2}N/dp_{T}d\eta$ is the differential yield per event in
collisions for a given centrality class and $\langle N_{\rm{bin}}
\rangle\,$ is the mean numbers of binary collisions corresponding to
this centrality.  Using a Monte Carlo Glauber calculation, as
described in~\cite{STARdAu200}, the mean number of binary collisions
for the 0-20\% and 40-100\% centrality classes were determined to be
15.0$\pm$1.1 and 4.0$\pm$0.3, respectively.  Fig.~\ref{Rcp} shows the
ratio of the central to peripheral spectra in d+Au collisions for various
pseudorapidity regions.  The error bars on each distribution are the
quadratic sum of statistical and systematic uncertainties; the latter
are due to uncertainties in our background subtraction technique.  An
overall error of about 10\% due to the uncertainty in normalization is
indicated by the band on the left portion of the figure.  The $R_{CP}$
in Au+Au collisions at $\sqrt{s_{NN}}=200$ GeV~\cite{RcpAuAu200} is
shown on the bottom of the plot.
$R_{CP}^{\rm{dAu}}$ distributions for each pseudorapidity selection exhibit
a rise with increasing \pT, exceeding unity at \pT$\sim$1-2 GeV/$c$.
At low \pT, the $R_{CP}^{\rm{dAu}}$ distribution is highest for the
most backward pseudorapidity region and systematically decreases the
more forward in pseudorapidity the ratio is taken.  The trend in the
pseudorapidity dependence indicates that the Cronin effect is more
pronounced in the gold hemisphere of the collision, consistent with
the measured asymmetry between backward and forward rapidity.  Our
measurement of $R_{CP}^{\rm{dAu}}$ shows no significant suppression at
\pT~of 2-6 GeV/$c$.  This result stands in contrast to the Au+Au measurements, where
$R_{CP}$ was observed to be well below unity for \pT~$<$ 12 GeV/$c$.
The results for $R_{CP}^{\rm{dAu}}$ are consistent
with calculations in pQCD models incorporating both Cronin
enhancement and nuclear shadowing
~\cite{oldwang,zhang,Kopeliovich,Accardi,Vitev}. However, the models
based on incoherent parton scattering at the initial stage fail to reproduce the rapidity
dependence in both backward-to-forward ratios and $R_{CP}^{\rm{dAu}}$.

In summary, we have studied the centrality and pseudorapidity
dependence of charged hadron production in d+Au collisions at
$\sqrt{s_{NN}}=200$ GeV.  The inclusive charged hadron multiplicity is
observed to be higher in the gold hemisphere than the deuteron
hemisphere of the collision.  The gluon saturation, HIJING, and AMPT
models cannot be ruled out from the integrated charged particle
pseudorapidity distributions. Ratios of backward-to-forward
pseudorapidity transverse momentum distributions are above unity for \pT~below 5
GeV/$c$.  Our measurement of $R_{CP}^{\rm{dAu}}$ shows no suppression
at \pT~of 2-6 GeV/$c$, with the ratio taken at backward
pseudorapidities being slightly higher than at forward pseudorapidities.
The incoherent multiple scattering of partons in the initial state alone
cannot reproduce the observed pseudorapidity asymmetry, while
the latest calculations in a gluon saturation model stand in qualitative agreement with
our observations.

We are grateful to D. Kharzeev and K. Tuchin for providing us their
saturation results and X.N. Wang for valuable discussions and for
providing us with the multiple partonic scattering results.
We thank the RHIC Operations Group and RCF at BNL, and the
NERSC Center at LBNL for their support. This work was supported
in part by the HENP Divisions of the Office of Science of the U.S.
DOE; the U.S. NSF; the BMBF of Germany; IN2P3, RA, RPL, and
EMN of France; EPSRC of the United Kingdom; FAPESP of Brazil;
the Russian Ministry of Science and Technology; the Ministry of
Education and the NNSFC of China; Grant Agency of the Czech Republic,
FOM and UU of the Netherlands, DAE, DST, and CSIR of the Government
of India; Swiss NSF; and the Polish State Committee for Scientific 
Research.


\begin{thebibliography}{99}
\bibitem{dima} D. Kharzeev, Nucl. Phys. A $\bf{730}$, 448 (2004).
\bibitem{cgc} D. Kharzeev {\it et al.}, Phys. Lett. B $\bf{561}$, 93 (2003);
  J. Jalilian-Marian {\it et al.}, Phys. Lett. B $\bf{577}$, 54 (2003);
  J.L. Albacete {\it et al.}, Phys. Rev. Lett. $\bf{92}$, 082001
  (2004); D. Kharzeev {\it et al.}, Phys. Rev. D $\bf{68}$, 094013 (2003);
  R. Baier {\it et al.}, Phys. Rev. D $\bf{68}$, 054009 (2003). 
\bibitem{wang} X.N. Wang, Phys.Lett. B$\bf{565}$, 116 (2003).
\bibitem{ziwei} Z.W. Lin and C.M. Ko, Phys.Rev. C$\bf{68}$, 054904 (2003).
\bibitem{Cronin} J.W. Cronin {\it et al.}, Phys. Rev. D $\bf{11}$, 3105 (1975).
\bibitem{STARdAu200} J. Adams {\it et al.}, Phys. Rev. Lett. $\bf{91}$, 072304 (2003).
\bibitem{PHENIXdAu200} S.S. Adler {\it et al.}, Phys. Rev. Lett. $\bf{91}$, 072303 (2003). 
\bibitem{PHOBOSdAu200} B.B. Back {\it et al.}, Phys. Rev. Lett. $\bf{91}$, 072302 (2003).
\bibitem{BRAHMSdAu200} I. Arsene {\it et al.}, Phys. Rev. Lett. $\bf{91}$, 072305 (2003).
\bibitem{starpiddAu} J. Adams {\it et al.}, nucl-ex/0309012.
\bibitem{strikman} L. Frankfurt, V. Guzey and M. Strikman, hep-ph/0303022.
\bibitem{vogt} R. Vogt, hep-ph/0405060.
\bibitem{KKT2} D. Kharzeev, Y.V. Kovchegov and K. Tuchin, hep-ph/0405045; D. Kharzeev and K. Tuchin, private communication.
\bibitem{finalstate} R.C. Hwa and C.B. Yang, nucl-th/0403001.; R.C. Hwa and C.B. Yang, Phys. Rev. C $\bf{70}$, 037901 (2004).;  R.C. Hwa, private communication.
\bibitem{Eloss1} M. Gyulassy and M. Plumer, Phys. Lett. B $\bf{243}$, 432 (1990).
\bibitem{Eloss2} X.N. Wang and M. Gyulassy, Phys. Rev. Lett. $\bf{68}$, 1480 (1992).
\bibitem{KLM} D. Kharzeev, E. Levin, and L. McLerran, Phys. Lett. B $\bf{561}$, 93 (2003).
\bibitem{shadow}  K.J. Eskola and H. Honkanen, Nucl. Phys. A $\bf{713}$, 167 (2003).
\bibitem{TPC} M. Anderson {\it et al.}, Nucl. Instrum. Meth. A $\bf{499}$, 659 (2003).
\bibitem{ZDC} C. Adler {\it et al.}, Nucl. Instrum. Meth. A $\bf{461}$, 337 (2001).
\bibitem{FTPC} K.H. Ackermann {\it et al.}, Nucl. Instrum. Meth. A $\bf{499}$, 713 (2003).
\bibitem{HIJING} X.N. Wang and M. Gyulassy, Phys. Rev. D $\bf{44}$, 3501 (1991). Version 1.382 is used.
\bibitem{HIJINGshadow} S.Y. Li and X.N. Wang, Phys. Lett. B $\bf{527}$, 85 (2002).
\bibitem{EKS} K.J. Eskola, V.J. Kolhinen and C.A. Salgado, Eur. Phys. J. C$\bf{9}$, 61 (1999).
\bibitem{Accardi} A. Accardi, hep-ph 0212148, and references therein.
\bibitem{KKT1} D. Kharzeev, Y.V. Kovchegov and K. Tuchin, Phys. Rev. D $\bf{68}$, 094013 (2003).
\bibitem{Qiu-Vitev} J. Qiu and I. Vitev, hep-ph/0405068.
\bibitem{BRAHMS} I. Arsene {\it et al.}, nucl-ex/0403005.
\bibitem{paulRcp} J. Adams {\it et al.}, Phys. Rev. Lett. $\bf{92}$, 052302 (2004). 
\bibitem{reco1} Z.W. Lin and C.M. Ko, Phys. Rev. Lett. $\bf{89}$, 202302 (2002).
\bibitem{reco2} D. Molnar and S.A. Voloshin, Phys. Rev. Lett. $\bf{91}$, 092301 (2003).; R.J. Fries {\it et al.}, Phys. Rev. Lett. 90, 202303 (2003).
\bibitem{reco3} R.C. Hwa and C.B. Yang, nucl-th/0401001.
\bibitem{reco4} V. Greco {\it et al.}, Phys. Rev. Lett. $\bf{90}$, 202302 (2003).
\bibitem{RcpAuAu200} J. Adams {\it et al.}, Phys. Rev. Lett. $\bf{91}$, 172302 (2003).
\bibitem{oldwang} X.N. Wang, Phys. Rev. C $\bf{61}$, 064910 (2000);Phys. Lett. B $\bf{565}$, 116 (2003).
\bibitem{zhang} Y. Zhang {\it et al.}, Phys. Rev. C $\bf{65}$, 034903 (2002).
\bibitem{Kopeliovich} B.Z. Kopeliovich {\it et al.}, Phys. Rev. Lett. $\bf{88}$, 232303 (2002).
\bibitem{Vitev} I. Vitev, Phys. Lett. B $\bf{562}$, 36 (2003).
\end{thebibliography}
\end{document}